\documentclass[10pt,letterpaper]{article}
\usepackage{opex3}
\usepackage{ulem}
\usepackage{graphicx}
\usepackage{color}

\newcommand{\ket}[1]{\ensuremath{\left|#1\right\rangle}}

\begin{document}

\title{Generation of polarization-entangled photon pairs in a Bragg reflection waveguide}

\author{A. Vall\'{e}s,$^{1,*}$ M. Hendrych,$^1$ J. Svozil\'{i}k,$^{1,2}$ R. Machulka,$^2$ P. Abolghasem,$^3$ D. Kang,$^3$ B. J.
Bijlani,$^3$ A. S. Helmy,$^3$ and J. P. Torres$^{1,4}$}
\address{$^1$ICFO-Institut de Ci{\`{e}}ncies Fot{\`{o}}niques, Mediterranean
Technology Park, Av. Carl Friedrich Gauss 3, 08860 Castelldefels,
Barcelona, Spain \\
$^2$ RCPTM, Joint Laboratory of Optics PU and IP AS CR, 17. listopadu 12, 771 46 Olomouc, Czech Republic \\
$^3$ Edward S. Rodgers Department of Electrical and Computer
Engineering, University of Toronto, 10 Kings College road,
Toronto, Ontario M5S3G4, Canada \\
$^4$ Department of Signal Theory and Communications, Universitat
Polit{\`{e}}cnica de Catalunya, Jordi Girona 1-3, Campus Nord D3,
08034 Barcelona, Spain}

\email{adam.valles@icfo.es}

\begin{abstract}
We demonstrate experimentally that spontaneous parametric
down-conversion in an Al$_x$Ga$_{1-x}$As semiconductor Bragg
reflection waveguide can make for paired photons highly entangled
in the polarization degree of freedom at the telecommunication
wavelength of 1550 nm. The pairs of photons show visibility higher
than $90\%$ in several polarization bases and violate a
Clauser-Horne-Shimony-Holt Bell-like inequality by more than $3$
standard deviations. This represents a significant step toward the
realization of efficient and versatile self pumped sources of
entangled photon pairs on-chip.
\end{abstract}


\section{Introduction}
Entanglement is not only a fundamental concept in Quantum
Mechanics with profound implications, but also a basic ingredient
of many recent technological applications that has been put
forward in quantum communications and quantum computing
\cite{nielsen2000,zeilinger2000}. Entanglement is a very special
type of correlation between particles that can exist in spite of
how distant they are. Nevertheless, the term entanglement is
sometimes also used to refer to certain correlations existing
between different degrees of freedom of a single particle
\cite{nagali2009}.
\par By and large, the most common method to generate photonic
entanglement, that is entanglement between photons, is the process
of spontaneous parametric down-conversion (SPDC)
\cite{torres2011}. In SPDC, two lower-frequency photons are
generated when an intense higher-frequency pump beam interacts
with the atoms of a non-centrosymetric nonlinear crystal.
Entanglement can reside in any of the degrees of freedoms that
characterize light: angular momentum (polarization and orbital
angular momentum), momentum and frequency, or in several of them,
what is known as hyper-entanglement. Undoubtedly, polarization is
the most widely used resource to generate entanglement between
photons thanks to the existence of many optical elements to
control the polarization of light and to the easiness of its
manipulation when compared to other characteristics of a light
beam, e.g., its spatial shape or bandwidth.
\par The implementation of entanglement-based photonic technologies
should consider the development of high-efficient, compact, and
highly tunable sources of entangled photons. High efficiency helps
to reduce the pump power required to generate a high flux of
down-converted photons, and broad tunability allows the
preparation of different types of quantum states. Compactness
makes possible to use the entanglement source under a greater
variety of circumstances, such as, for instance, would be the case
in free space applications \cite{fabian2012}. Along these lines,
the use of waveguides is very advantageous. Contrary to the case
of SPDC in bulk crystals, where a very large number of spatial
modes is generated, and only a few of them effectively contribute
to the generated entangled state, the use of waveguides allows the
reduction of the number of modes to a few guided modes
\cite{banaszek2001}, and, in this way, it contributes to enhance
the overall efficiency of the nonlinear interaction
\cite{fiorentino2007}.
\par The capability of integration of the SPDC source with other
elements, such as the pumping laser or optical circuits, in a
single platform, might be crucial for the implementation of
entanglement-based quantum circuits in an out-of-the-lab
environment. Semiconductor technologies are nowadays a mature
technology that offers a myriad of possibilities, and that allows
the fabrication of an integrated monolithic source of entangled
photon pairs. Bragg reflection waveguides (BRWs) in
Al$_x$Ga$_{1-x}$As could make possible the integration of all of
these elements in a single semiconductor platform.
In the last few years, different nonlinear optics processes have
been observed experimentally in Al$_x$Ga$_{1-x}$As BRWs, such as
second-harmonic generation~\cite{helmy2007,abolghasem2009a},
difference-frequency generation~\cite{han2010} and spontaneous
parametric down-conversion~\cite{horn2012}. Also, BRWs have been
demonstrated as edge-emitting diode lasers where the fundamental
lasing mode is a photonic bandgap mode or a Bragg
mode~\cite{bijlani2009}, and electrically pumped parametric
fluorescence was demonstrated subsequently
\cite{bijlani_CLEO_2011}.

GaAs based waveguides show a broad transparency window
($1-17~\mu$m), large damage threshold, low linear propagation loss
and an extremely high non-linear
coefficient~\cite{abolghasem2012}. In spite of GaAs being an
isotropic material, not showing birefringence, phase-matching can
nevertheless be reached between high frequency light propagating
as a photonic bandgap mode, or Bragg mode, and low frequency light
beams propagating as bound modes based on total-internal
reflection (see Fig. \ref{fig1})~\cite{helmy2006}. Fortunately,
strong modal dispersion in BRWs offers significant control over
the spectral width~\cite{abolghasem2009b} and the type of spectral
correlations \cite{svozilik2011} of the emitted photons.
\par In this paper, we demonstrate that the use of BRWs allows the
generation of highly entangled pairs of photons in polarization
via the observation of the violation of the
Clauser-Horne-Shimony-Holt (CHSH) Bell-like
inequality~\cite{CHSH1969}. Bell's inequalities are a way to
demonstrate entanglement~\cite{gisin1991}, since the violation of
a Bell's inequality makes impossible the existence of one joint
distribution for all observables of the experiment, returning the
measured experimental probabilities~\cite{fine1982}.

\par In a previous work~\cite{horn2012}, the existence of
time-correlated paired photons generated by means of SPDC in BRWs
was reported, but the existence, and quality, of the entanglement
present was never explored. The generation of polarization
entanglement in alternative semiconductor platforms has been
demonstrated recently in a silicon-based wire waveguide
\cite{matsuda2012}, making use of four-wave mixing, a different
nonlinear process to the one considered here, and in a AlGaAs
semiconductor waveguide~\cite{ducci2013}, where as a consequence
of the opposite propagation directions of the generated
down-converted photons, two type-II phase-matched processes can
occur simultaneously.
\section{Device description and SHG characterization}
\begin{figure}[t]
\centering \includegraphics[width=11cm]{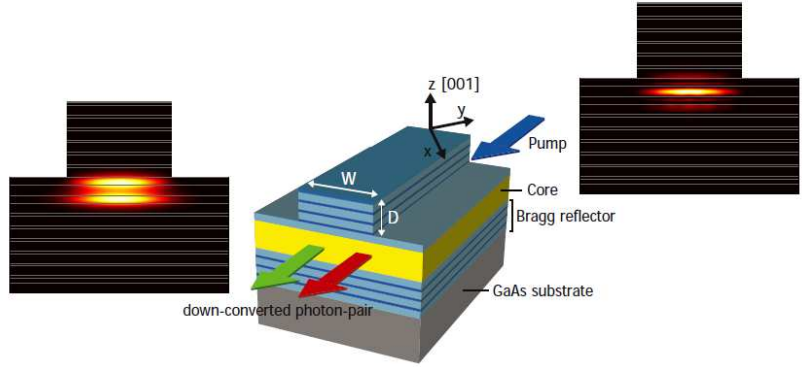} \caption{Bragg
reflection waveguide structure used to generate paired photons
correlated in time and polarization (type-II SPDC) at the
telecommunication window ($1550$ nm). The insets show the spatial
shape of the pump mode that propagates inside the waveguide as a
Bragg mode, and the spatial shape of the down-converted, which are
modes guided by total internal reflection (TIR). W: width of the
ridge; D: depth of the ridge.} \label{fig1}
\end{figure}

A schematic of the BRW used in the experiment is shown in
Fig.~\ref{fig1}. Grown on an undoped [001] GaAs substrate, the
epitaxial structure has a three-layer waveguide core consisting of
a 500 nm thick Al$_{0.61}$Ga$_{0.39}$As layer and a 375 nm
Al$_{0.20}$Ga$_{0.80}$As matching-layer on each side. These layers
are sandwiched by two symmetric Bragg reflectors, with each
consisting of six periods of 461 nm Al$_{0.70}$Ga$_{0.30}$As/129
nm Al$_{0.25}$Ga$_{0.65}$As. A detailed description of the
epitaxial structure can be found in~\cite{abolghasem2009c}. The
wafer was then dry etched along [110] direction to form ridge
waveguides with different ridge widths. The device under test has
a ridge width of $4.4~\mu$m, a depth of $3.6~\mu$m and a length of
1.2~mm. The structure supports three distinct phase-matching
schemes for SPDC, namely: type-I process where the pump is
TM-polarized and the down-converted photon pairs are both
TE-polarized; type-II process where the pump is TE-polarized while
the photons of a pair have mutually orthogonal polarization
states, and type-0 process where all three interacting photons are
TM-polarized~\cite{abolghasem2012}. For the experiment here, we
investigate type-II SPDC, which is the nonlinear process that
produce the polarizations of the down-converted photons required
to generate polarization entanglement. Since both photons show
orthogonal polarizations, after traversing a non-polarizing beam
splitter and introducing in advance an appropriate temporal delay
between them, they can result in a polarization-entangled pair of
photons.

During the fabrication process of the BRW, slight changes in the
thickness and aluminium concentration of each layer result in
small displacements of the actual phase-matching wavelength from
the design wavelength. For this reason, we first use second
harmonic generation (SHG) before examining SPDC to determine the
pump phase-matching wavelength for which the different schemes
(type-I, type-II or type-0) are more efficient.
\par
\begin{figure}[t]
\centering
\includegraphics[width=11cm]{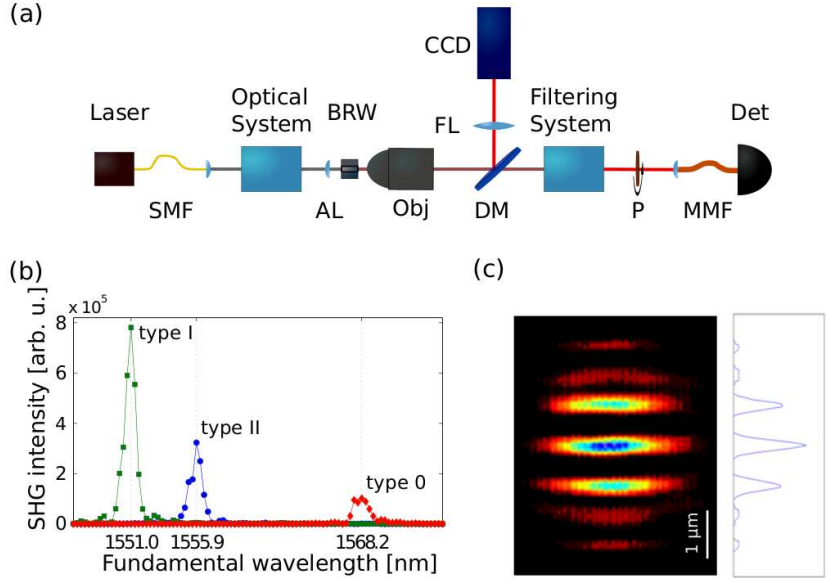}
\caption{(a) Experimental setup for SHG. The pump laser is a
tunable external-cavity semiconductor laser (TLK-L1550R,
Thorlabs). The Optical System consists of a linear power
attenuator, polarization beam splitter and a half-wave plate. The
Filtering System consists of a neutral density filter and low-pass
filter. SMF: single-mode fiber; AL: aspheric lens; BRW: Bragg
reflection waveguide; Obj: Nikon $50\times$; DM: dichroic mirror;
FL: Fourier lens; CCD: Retiga EXi Fast CCD camera; P: polarizer;
MMF: multi-mode fiber; Det: single-photon counting module (SPCM,
PerkinElmer). (b) Phase-matching curve of the BRW as a function of
the wavelength of the fundamental wave. (c) Beam profile of the
Bragg mode of the second harmonic wave generated by means of the
SHG process, captured with a CCD camera after imaging with a
magnification optical system of $100\times$ (Fourier lens with
focal length f=400 mm).} \label{fig2}
\end{figure}

 The experimental arrangement for SHG is shown in Fig.~\ref{fig2}(a). The
wavelength of a single-frequency tunable laser (the fundamental
beam) was tuned from $1545$~nm to $1575$~nm. An optical system
shapes the light into a Gaussian-like mode, which is coupled into
the BRW to generate the second harmonic beam by means of SHG. At
the output, the power of the second harmonic wave is measured to
determine the efficiency of the SHG process. Figure~\ref{fig2}(b)
shows the phase-matching tuning curve showing the dependency of
generated second-harmonic power on the fundamental wavelength.
From the figure, three resonance SH features could be resolved
corresponding to the three supported phase-matching schemes. As
mentioned earlier, the process of interest here is type-II. For
this particular type of phase-matching, maximum efficiency takes
place at the fundamental wavelength of $1555.9$~nm. To generate
the second harmonic beam by means of type-II SHG in
Fig.~\ref{fig2}(b), we use a half-wave plate to rotate the
polarization of the fundamental light coming from the laser by
$45$-degrees, to generate the required fundamental beams with
orthogonal polarizations.

In BRW, phase-matching takes place between different types of
guided modes which propagate with different longitudinal
wavevectors. The fundamental beam (around $1550$ nm) corresponds
to a total internal reflection (TIR) mode, and the second harmonic
beam (around $775$ nm) is a Bragg mode. The measured spatial
profile of this Bragg mode is shown in Fig.~\ref{fig2}(c).

\section{Experimental set-up for the generation of polarization entanglement}
\begin{figure}[t]
\centering
\includegraphics[width=11cm]{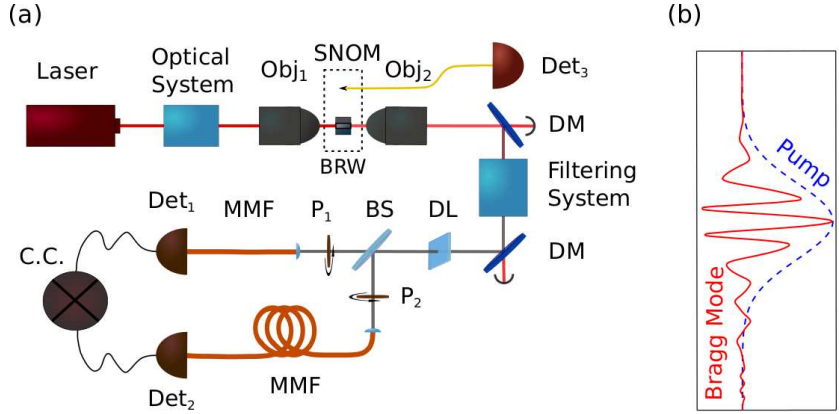}
\caption{(a) Experimental setup for SPDC. The Optical System is
composed of a linear power attenuator, spatial filter and beam
expander. SNOM: scanning near-field optical microscope probe; BRW:
Bragg reflection waveguide; Objectives: Obj$_1$ (Nikon
$100\times$) and Obj$_2$ (Nikon $50\times$); DM: dichroic mirror;
Filtering System: 2 DMs, band-pass and long-pass filters; DL:
delay line (birefringent plate); BS: beam splitter; P$_1$ and
P$_2$: linear film polarizers; MMF: multi-mode fiber; D$_1$ and
D$_2$: InGaAs single-photon counting detection modules; D$_3$:
low-power silicon detector; C.C.: coincidence-counting
electronics. (b) Amplitude profiles of the theoretical Bragg mode
and the Gaussian-like pump beam.} \label{fig3}
\end{figure}

The experimental setup used to generate polarization-entangled
paired photons and the measurement of the Bell-like inequality
violation is shown in Fig.~\ref{fig3}(a). The pump laser is a
tunable single-frequency diode laser with an external-cavity (DLX
110, Toptica Photonics) tuned to $777.95$ nm. Light from the laser
traverses an optical system, with an attenuator module, spatial
filter and beam expander, in order to obtain a proper input beam.
Even though the optimum option for exciting the pump Bragg mode
would be to couple directly into the photonic bandgap mode using a
spatial light modulator (SLM), the small feature size in the field
profile of the Bragg mode and its oscillating nature imposed
serious challenges for using an SLM. Therefore, we choose instead
to pump the waveguide with a tightly focused Gaussian pump beam
(see Fig. \ref{fig3}(b)) with a waist of $\sim 1.5~\mu$m, that is
coupled into the waveguide using a $100\times$ objective. Our
calculations show that the estimated modal overlap between the
Gaussian pump beam and the Bragg mode of the waveguide is around
$20\%$, which should be added to the total losses of the system.
A scanning near-field optical microscope (SNOM) probe was attached
to the support of the BRW, in order to perform sub-micrometric 3D
beam profile scans to maximize the coupling efficiency of the
incident pump beam into the pump Bragg mode. The power of the
laser light before the input objective was measured to be $13$ mW.
Taking into account the transmissivity of the objective for
infrared light~(70\%), the transmissivity of the facet of the BRW
(73\%) and the calculated overlap between the laser light and the
Bragg mode of the waveguide (around 20\%), the estimated pump
power available for SPDC process inside the waveguide is $\sim
1.3$ mW.
\par The generated down-converted photons are collected using
a $50\times$ objective and separated from the pump photons using
four dichroic mirrors (DM), band-pass and long-pass filters. Each
DM has a 99\% transmissivity at the pump wavelength. The
attenuation of the band-pass filter (45 nm FWHM bandwidth centered
at $1550$ nm) is $10^{-4}$, and the long-pass filter (cut-on
wavelength: $1500$ nm) introduces an additional attenuation of
$10^{-3}$ at the pump wavelength.

In general, photons propagating in a waveguide with orthogonal
polarizations have different group velocities (group velocity
mismatch, GVM), which in conjunction with non-negligible group
velocity dispersion (GVD), result in different spectra for the
cross-polarized photons \cite{Zhukovsky_PRA_2012}. As a
consequence, the polarization and frequency properties of the
photons are mixed. The two photons of a pair could be, in
principle, distinguished by their time of arrival at the
detectors, as well as their spectra, which diminishes the quality
of polarization entanglement achievable. In order to obtain
high-quality polarization entanglement, it is thus necessary to
remove all the distinguishing information coming from the
temporal/frequency degree of freedom. For this reason, the 45 nm
band pass filter was applied to remove most of the distinguishing
spectral information, and off-chip compensation was implemented
with a delay line to remove arrival time information.

A quartz birefringent plate with a length of~$1$ mm, vertically
tilted around 30$^\circ$ was used to introduce a $32$~fs time
delay between photons, which is experimentally found to be the
optimum value to erase temporal distinguishing information caused
by the group velocity mismatch (GVM) and the GVD.  The calculated
group velocities for TE and TM down-converted photons are
8.98$\times 10^7$ m/s and 9.01$\times10^7$ m/s, respectively. The
GVD parameter is $D \sim$ -7.9$\times10^2$ ps/(nm$\cdot$km) for
both polarizations. When considering these values of the GVM and
GVD, our calculations show that the optimum delay for generating
the highest degree of polarization entanglement is $\sim 31.2$ fs,
which agrees with the value obtained experimentally.

The down-converted photons are separated into arms $1$ and $2$
with a 50/50 beam splitter (BS) in order to generate a
polarization-entangled two-photon state of the form
\begin{equation}
 \ket{\Psi^{+}}=\frac{1}{\sqrt{2}}\left\{ \ket{H}_1 \ket{V}_2 + \ket{V}_1 \ket{H}_2 \right\},
\label{entangledState}
\end{equation}
where $\ket{H}$ and $\ket{V}$ denote the two possible
polarizations of the photons (horizontal and vertical),
propagating in arms $1$ or $2$. Horizontal (vertical) photons
corresponds to photons propagating inside the waveguide as TE (TM)
mode. We neglect cases where both photons leave the BS through the
same output port, by measuring only coincidences between photons
propagating in arms $1$ and $2$ (post-selection), which implies
that 50$\%$ of the generated pairs are not considered. Finally, to
measure Bell's inequality violations, the entangled photons are
projected into different polarization states with linear film
polarizers, and coupled into multi-mode fibers connected to InGaAs
single-photon detection modules (id201, idQuantique), where
optical and electronic delays are introduced to measure
coincidental events with time-to-amplitude converter (TAC)
electronics. The coincidences window for all measurements was set
to $3$ ns.
\section{Violation of the CHSH inequality}
To obtain a first indication that the pairs of photons propagating in arms $1$ and $2$ are truly entangled in the
polarization degree of freedom, so that their quantum state can be written of the form given by
Eq.~(\ref{entangledState}), one detects one of the photons, i.e., the photon propagating in arm $1$, after projection
into a specific polarization state $\ket{\Psi}_1=\cos \theta_1 \ket{H}_1 - \sin \theta_1 \ket{V}_1$
\cite{polarization_state}, and measures in coincidence the remaining photon after projection into a set of polarization
bases $\ket{\Psi}_2=\cos \theta_2 \ket{V}_2 + \sin \theta_2 \ket{H}_2$, with $\theta_2$ spanning from $0$ to $2\pi$
\cite{kwiat1995}. Ideally, the coincidence counts as a function of $\theta_2$ should follow the form of
$\cos^2(\theta_1+\theta_2)$, which yields a visibility $V=(Max-Min)/(Max+Min)$ of $100\%$. Therefore, the highest the
visibility measured, the highest the quality of the generated polarization-entangled state.
\par
\begin{figure}[t]
\centering
\includegraphics[width=11cm]{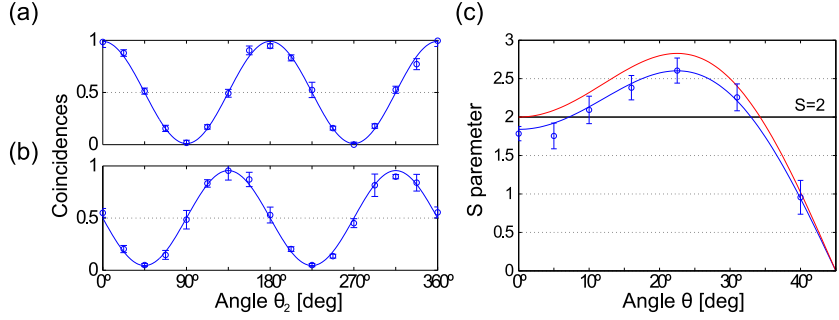}
\caption{Normalized coincidence measurements as a function of the
polarization state of photon $2$ when photon $1$ is projected into
a polarization state with: (a) $\theta_1=0^\circ$ and (b)
$\theta_1=45^\circ$. The data shown in (a) and (b) is subtracting
from the raw data the number of accidental coincidences. (c)
Violation of the CHSH inequality. Parameter $S$ as a function of
the angle $\theta$. The small blue circles with error bars
represent the experimental data with their standard deviations.
The blue solid curves in (a) and (b) are theoretical predictions
assuming that the visibility is $98\%$ in (a) and $91\%$ in (b).
The red (upper)  curve in (c) is the theoretical prediction for
$S$. The blue curve in (c) is the best fit. The inequality holds
if $S\le 2$. The maximum value attained is $S=2.61\pm 0.16$. The
data shown in (c) is without subtraction of accidental
coincidences. } \label{fig4}
\end{figure}

 Figures~\ref{fig4}(a) and (b) show the results of the measurements
for two specific cases: $\theta_1=0^{\circ}$ and
$\theta_1=45^{\circ}$. The measured visibility, subtracting the
accidental coincidences, is 98\% for $\theta_1=0^{\circ}$, and to
91\% for $\theta_1=45^{\circ}$. Without subtraction of accidental
coincidences, the corresponding measured visibility is 80\% for
$\theta_1=0^{\circ}$ and 77\% for $\theta_1=45^{\circ}$. The
accidental coincidences, with respect to the total number of
events counted, were measured experimentally, introducing an
electronic delay in the trigger of the second detector driving it
out of the detection window of the first detector. The same
electronic delay had to be introduced before the TAC electronics
in order to have the coincidence events from the same amount of
single events, but totally uncorrelated in this case. This
technique made possible to measure the correct visibility of the
fringes using the maximum efficiency detector settings, in order
to obtain lower standard deviation of the measurements. The
optimum trigger rate for this experiment was found to be
$100$~KHz, measuring an average of $3,550$ and $6,200$ photon
counts per second in each detector, and a maximum flux rate of
coincidences of $3$ pairs of photons per second. The low trigger
rate is one of the reasons for the observation of such a low flux
rate of down-converted photons observed, since it implies that the
detectors are closed most of the time. The detection window for
these measurements was set to $100$ ns.

In a CHSH inequality experiment \cite{CHSH1969}, one measures
photon coincidences between photon $1$, after being projected into
a polarization state defined by angles $\theta_1$ or
$\theta_1^{\prime}$, and photon $2$, after a similar polarization
projection defined by angles $\theta_2$ or $\theta_2^{\prime}$.
The CHSH inequality holds if
\begin{equation}
 S = | E(\theta_1, \theta_2) - E(\theta_1, \theta_2^{'}) + E(\theta_1^{'}, \theta_2) + E(\theta_1^{'},
\theta_2^{'})| \leq 2, \label{CHSH}
\end{equation}
where
\begin{equation}
E(\theta_1, \theta_2) = \frac{C(\theta_1, \theta_2) +
C(\theta_1^{\perp}, \theta_2^{\perp}) - C(\theta_1^{\perp},
\theta_2) - C(\theta_1, \theta_2^{\perp})}{C(\theta_1, \theta_2) +
C(\theta_1^{\perp}, \theta_2^{\perp}) + C(\theta_1^{\perp},
\theta_2) + C(\theta_1, \theta_2^{\perp})} \label{coincFrac}
\end{equation}
and $\theta_{1,2}^{\perp}=\theta_{1,2} + 90^{\circ}$.
Figure~\ref{fig4}(c) shows the value of the parameter $S$ as a
function of the angle $\theta$, where $\theta \equiv \theta_2 -
\theta_1 = \theta_2^{'} + \theta_1^{'} = - \theta_2 -
\theta_1^{'}$, which attains the maximum possible violation, i.e.,
$S=2\sqrt{2}$. For the ideal case, one would obtain $S(\theta) =
3\cos 2\theta - \cos 6\theta$, which is the red (upper) curve
depicted in Fig. \ref{fig4}(c). Sixteen measurements were
performed for each value of the angle $\theta$. For the maximum
inequality violation ($\theta=22.5^{\circ}$), the polarizer
settings were $\theta_1=0^\circ$, $\theta_1^{'}=-45^{\circ}$,
$\theta_2=22.5^{\circ}$ and $\theta_2^{\prime}=67.5^{\circ}$. In
this case, we obtained a value of the inequality of $S = 2.61\pm
0.16$, which represents a violation by more than $3$ standard
deviations. This represents a stronger violation of the CHSH
inequality than previously reported \cite{ducci2013} for a
vertically pumped BRW structure, where the measured value was
$S=2.23\pm0.11$.

Regarding the measurements of the $S$ parameter, no accidental
coincidences were subtracted from the absolute measurement
obtained. In order to increase the signal-to-noise ratio, the
detection window in both detectors was decreased to 20\% of its
previous time duration (from $100$ ns to $20$ ns), having thus a
corresponding decrease in total number of single and coincidence
counts detected. Now, the measured average flux rate is 600 and
500 photon counts per second in each detector, and a maximum value
of $0.3$ pairs of photons per second.

To estimate the efficiency of the SPDC process, we take into
account that the detection window is $\tau=20$ ns, and the trigger
rate of detection is $100$ kHz. The efficiency of each
single-photon detector is $25\%$. The pump power injected into the
BRW waveguide is estimated to be around $1.3$ mW. Assuming that
the transmissivity of each optical system, traversed by signal/
idler photons, not including detection efficiency, is $\sim 10\%$,
it results in an estimated SPDC efficiency of $\sim 10^{-10}$ in
the filtering bandwidth.

\section{Conclusions}
We have demonstrated that polarization-entangled paired photons
generated in a semiconductor Bragg reflection waveguide (BRW) show
a visibility higher than $90\%$ in all the bases measured, a
requisite for obtaining high quality entanglement. It has also
been experimentally demonstrated that the generated two-photon
state clearly violate the CHSH inequality, and that the presented
BRW source can be considered an expedient source of high-quality
polarization-entangled two-photon states.

Alternative BRW configurations with no need of post-selection of
down-converted photons can be implemented using, for instance,
non-degenerate SPDC, where signal and idler photons bear different
wavelengths \cite{Svozilik_OE_2012}, or concurrency of two
conversion processes \cite{Kang_OL_2012}. For this, one can make
use of the great versatility offered by BRWs and design the layer
structure to achieve phase-matching at the required wavelengths.
Optimization of the generation rate of down-converted photons can
be achieved by optimizing the layer thicknesses and Al
concentrations, so that the mode overlap between photons at
different wavelength increases. BRW made of AlGaAs can potentially
offer higher generation rates than ferroelectrical waveguides made
of PPLN or PPKTP, since they show a much higher second-order
nonlinear coefficient. However, in practice, both the pump and the
down-converted modes are subject to losses, chiefly by two
processes: radiation losses, mainly in the Bragg modes, and
scattering of light due to surface roughness
\cite{Zhukovsky_JOSAB_2012}. Fortunately, improvements in design
and fabrication of the BRW could reduce the losses of the pump and
down-converted waves, increase mode overlap and enhance the
coupling efficiency of the pump light into the pump mode that
propagates in the waveguide.

It is important to note that the platform described and used here
offers the unique possibility of integrating the pump laser with
the nonlinear element to enable self-pumped on-chip generation of
polarization entanglement, without the use of off-chip
compensation and bandpass filtering, as is carried in this work.
There are two theoretical proposals to achieve this aim, both use
dispersion engineering of the BRWs. One uses type-II process in a
BRW with zero-GVM \cite{Zhukovsky_PRA_2012}, while the other one
uses concurrent type-I and type-0 processes \cite{Kang_OL_2012}.

In combination with the development of quantum circuits composed
of properly engineered arrays of waveguides, and the integration
of the laser pump source in the same chip, our results show that
semiconductor technology based on the use of BRW in
Al$_{x}$Ga$_{1-x}$As is a promising path to develop integrated
entanglement-based quantum circuits.

\section*{Acknowledgments}
We would like to thank M. Micuda for his collaboration in certain
stages of the experiment and R. de J. Le\'{o}n-Montiel for useful
discussions. This work was supported by Projects FIS2010-14831 and
FET-Open 255914 (PHORBITECH). J. S. thanks the project FI-DGR 2011
of the Catalan Government. This work was also supported in part by
projects CZ.1.05/2.1.00/03.0058 of the Ministry of Education,
Youth and Sports of the Czech Republic and by PrF-2012-003 of
Palack\'{y} University. P. Abolghasem, D. Kang and A. S. Helmy
acknowledge the support of Natural Sciences and Engineering
Research Council of Canada (NSERC) for funding this research and
CMC Microsystems for growing the wafer. M. Hendrych is currently
with Radiantis.
\end{document}